# The Structural Fate of Individual Multicomponent Metal-Oxide Nanoparticles in Polymer Nanoreactors

Jingshan S. Du[+], Peng-Cheng Chen[+], Brian Meckes, Zhuang Xie, Jinghan Zhu, Yuan Liu, Vinayak P. Dravid,* and Chad A. Mirkin*

**Abstract:** Multicomponent nanoparticles can be synthesized with either homogeneous or phase-segregated architectures depending on the synthesis conditions and elements incorporated. To understand the parameters that determine their structural fate, multicomponent metal-oxide nanoparticles consisting of combinations of Co, Ni, and Cu were synthesized via scanning probe block copolymer lithography and characterized using correlated electron microscopy. These studies revealed that the miscibility, ratio of the metallic components, and the synthesis temperature determine the crystal structure and architecture of the nanoparticles. A Co-Ni-O system forms a rock salt structure largely due to the miscibility of CoO and NiO, while Cu-Ni-O, which has large miscibility gaps, forms either homogeneous oxides, heterojunctions, or alloys depending on the annealing temperature and composition. Moreover, a higher ordered structure, Co-Ni-Cu-O, was found to follow the behavior of lower ordered systems.

Oxide nanoparticles (NPs) consisting of 3d transition metal elements have unique physical and chemical properties, and constitute a particularly important class of functional materials.[1] In order to further expand the functionality of oxide NPs, one can integrate multiple components into these structures to create homogeneous and phase-segregated mixed oxides. By controlling the composition, crystal structure, and organization of interfaces and defects, one can control their band structure and surface chemical properties.[2] Therefore, synthesizing well-defined, multicomponent 3d transition metal oxide NPs and related heterostructures has increasing importance in many fields, including (photo)(electro)catalysis,[3] optics, optoelectronics,[4] magnetics,[5] and sensing.[6] However, how the elemental composition and synthesis conditions determine the crystal structure and architecture of multicomponent and heterostructured NPs is poorly understood. This lack of understanding has prevented a more thorough evaluation of particle structure-function relationships. To overcome this challenge, one needs: 1) the ability to synthesize and monitor individual particles within a localized nanoscopic reactor, and 2) a high throughput approach to both synthesis and screening.

Scanning probe block copolymer lithography (SPBCL) is a synthetic technique that allows one to address these challenges. SPBCL enables the synthesis of individual metal,[7] oxide,[7b] and sulfide[8] NPs at predefined locations on a substrate. Unlike other techniques that enable localized delivery of NPs, such as sol-gel deposition of oxide nanostructures via dip-pen nanolithography (DPN),[9] or printing of preformed NPs,[10] with SPBCL, metal-coordinated block copolymers are deposited onto a substrate in the form of attoliter hemispherical domes (polymer nanoreactors). Each reactor, when heated under the appropriate environment, results in the formation of a single NP.[7] Significantly, this process allows one to directly control the stoichiometry and thermal processing while synthesizing on substrates compatible with electron microscopy. These properties make SPBCL a powerful tool for synthesizing NPs with precise control over composition and size, making it an ideal platform for the systematic study of crystal structure and elemental composition and distribution with correlated electron microscopy. Moreover, since SPBCL allows one to synthesize and study many particles in parallel, NP formation can be studied in a combinatorial and statistically meaningful fashion.[11]

Although SPBCL has been used to study oxide NPs containing a single metal element,[7b] it has not been extended to multicomponent metal oxide formation. To reliably make a given multicomponent particle with a relatively well-defined structure and stoichiometry, the synthesis of each particle must be studied over a range of temperatures and compositions. In the case of polyelemental systems, we also need to distinguish mixed, homogeneous systems from phase-segregated ones. Herein, we have used SPBCL to synthesize a series of metal-oxide NPs in a homogeneous mixed (Co, Ni) and a heterogeneous phase-segregated state (Cu, Ni), respectively. To accomplish this goal, combinations of bivalent Co, Ni, or Cu nitrate salt precursors were dissolved in an aqueous solution of the block copolymer poly(ethylene oxide)-*block*-poly(2-vinylpyridine) (PEO-*b*-P2VP), to create inks with a pyridyl group to metal ion ratio of 32:1. These polymer inks were coated on DPN tips and then deposited on a silicon nitride thin film substrate as hemispherical nanoreactors. Subsequent thermal annealing under Ar gas converted the precursors into individual NPs (Figure 1A). Importantly, thermal decomposition of the nitrate salts at moderate temperatures (300 500 °C) results in the formation of metal-oxides during this process, while limiting diffusion between the substrate and NPs, which occurs at higher temperatures. These NPs were spatially distributed hundreds of nanometers to microns apart at pre-defined locations, thus enabling correlated and repeated characterization using both diffraction-based and analytical electron microscopy on the same particle.

[*]  J. S. Du,[+] P.-C. Chen,[+] J. Zhu, Prof. V. P. Dravid, Prof. C. A. Mirkin
Department of Materials Science and Engineering
Northwestern University
Evanston, IL 60208 (USA)
E-mail: chadnano@northwestern.edu (C.A.M.)
         v-dravid@northwestern.edu (V.P.D.)

Dr. B. Meckes, Dr. Z. Xie, Y. Liu, Prof. C. A. Mirkin
Department of Chemistry
Northwestern University
Evanston, IL 60208 (USA)

J. S. Du, P.-C. Chen, Dr. B. Meckes, Dr. Z. Xie, J. Zhu, Y. Liu, Prof. V. P. Dravid, Prof. C. A. Mirkin
International Institute for Nanotechnology
Northwestern University
Evanston, IL 60208 (USA)

[+] These authors contributed equally to this work.

Supporting information for this article can be found under:

[Figure 1, double column]

We began our study by initially evaluating a fully miscible oxide system, Co-Ni-O, in the context of a bulk thin film structure. X-ray diffraction (XRD) shows that within thin films, both metal elements form oxides with a rock salt structure (CoO and NiO, respectively), upon thermal annealing at 500 °C (Figure 1B and Supporting Information, Figure S1). Previous reports suggest that bulk samples consisting of mixtures of CoO and NiO form a homogeneous rock salt phase when annealed above 760 °C,[12] while low-temperature wet chemical synthetic procedures result in multi-crystalline (Co,Ni)O nanostructures.[13] Since the critical temperatures of phase transitions in NPs are often depressed below those of bulk materials,[14] we hypothesize that Co-Ni-O NPs may be able to form a homogeneous rock salt phase at lower temperatures. By annealing polymer nanoreactors containing different molar ratios of Co and Ni precursors (9:1 to 1:9) at 500 °C, a series of mixed oxide NPs were generated. Energy-dispersive X-ray spectroscopy (EDS) shows that Co, Ni, and O co-exist in the individual NPs (Figure S2 and Figure S3). To examine the distribution of the metal elements within the Co-Ni-O NPs, EDS mapping in scanning transmission electron microscopy (STEM) mode was performed and correlated to the high angle annular dark field (HAADF) images. In support of our hypothesis that homogeneous particles will form at temperatures lower than bulk samples, the Co-Ni-O NPs exhibited a uniform distribution of the two metal elements (Figure 1C). The relationship between composition and crystal structure of each NP was further established by acquiring correlated high-resolution transmission electron microscopy (HRTEM) images, EDS spectra, and electron diffraction patterns of individual NPs (Figure 1D). The atomic ratio of Co to Ni for each NP was quantified based on the EDS spectra. Electron diffraction (Figure 1D, right) and the corresponding Fast Fourier Transform (FFT) for the HRTEM images (Figure S4) reveal that a single-crystalline Co-Ni-O NP forms a rock salt structure over the entire composition spectrum.

In contrast to a fully miscible system like Co-Ni-O that forms a rock salt structure, major oxides of Cu and Ni have large miscibility gaps in bulk due to their distinct crystal structures.[15] In addition, since copper oxides have lower decomposition energies,[16] they more readily thermally decompose into the metallic form at elevated temperatures compared to NiO. At 300 °C, we found that polymer nanoreactors containing the Cu precursor resulted in cuprite ($Cu_2O$) NPs (Figure 2A and Figure S5A). The characteristic {011} diffraction spots and derived patterns,[17] as simulated in Figure S6, were used to identify the cuprite structure of the NPs. When the annealing temperatures were above 300 °C, the cuprite decomposed to metallic Cu during our synthesis. When we synthesized larger particles, we observed intermediate NPs with partial $Cu_2O$ and Cu regions as determined by the presence of four crystal domains in HRTEM (Figure S7A). These domains were identified to be along cuprite [211] and face centered cubic (*fcc*)-Cu [011] zone axes, respectively, according to their FFT (Figure S7B). XRD also confirmed the coexistence of cuprite and *fcc*-Cu phases in the bulk sample at elevated temperatures (Figure S8). This temperature dependence for Cu and its oxide, along with the fact that NiO remains stable at 500 °C, suggests that multiple heterostructured NPs may be synthesized by annealing polymer nanoreactors containing both Cu and Ni nitrate precursors (Figure S9).

[Figure 2, 1 ½ column]

For this system, annealing at 300 °C under Ar generated Cu and Ni oxide-based NPs, as determined by EDS (Figure S10). At low Ni content, below ~15% (defined as the atomic fraction of Ni in all metal species), the cuprite structure was formed, as confirmed by the correlated HRTEM, EDS, and diffraction patterns (Figure 2B and Figure S5B).

With increasing Ni content (between ~15% and ~65%), we no longer observed homogeneous single-crystalline cuprite NPs. Instead, phase-segregated heterojunction NPs with both Cu-rich and Ni-rich regions were generated (Figure 2C, Figure S11 and Figure S12). According to the HAADF image and corresponding EDS mapping, the interfaces between these two types of regions are broad, suggesting relatively low interfacial energy between the two phases. To identify the crystal structures in these heterojunction NPs, correlated HRTEM was performed on the same NP. In Figure 2C, region (**1**) is a Cu-rich one according to EDS mapping. The related FFT clearly shows four pairs of reflections indicative of the cuprite [011] zone axis, including the characteristic {011} spots. The Ni-rich region, labeled (**2**) in Figure 2C, only shows one pair of reflections in the FFT, suggesting that it is not on a zone axis. By tilting the sample slightly, another lattice was observed (Figure S11A and B). This lattice, combined with the previous lattice, reveals a pattern along the rock salt [211] zone axis, which is close to that expected for NiO (Figure S11C). These observations strongly suggest the formation of cuprite-rock salt heterojunctions in the Cu-Ni-O NPs with intermediate metal ratios.

With greater Ni content (> ~65%), NPs with a homogeneous distribution of Cu and Ni were formed, as confirmed by EDS mapping (Figure 2D). Correlated HRTEM, EDS, and electron diffraction (Figure 2E) were used to identify a rock salt crystal structure. This structure is similar to that of the Co-Ni-O NPs that we first evaluated (Figure 1B) but with Co replaced with Cu.

[Figure 3, single column]

At an annealing temperature of 500 °C that results in thermal decomposition of cuprite, particles with low Ni content (< ~35%) result in metal alloy NPs of Cu and Ni without significant oxygen signals in the EDS spectra (Figure S13). EDS mapping and correlated HRTEM, EDS, and electron diffraction show that homogeneous, *fcc*-structured alloy NPs are formed (Figure 3A and B). This result is similar to previous observations for Cu-Ni alloy NPs formed under reductive annealing in $H_2$ gas,[11b] but it stands in contrast to our previous observation that Ni nitrates form stable NiO NPs when annealed in Ar *without* Cu present in the system. Therefore, the formation of a Cu-Ni alloy indicates that Cu formation may synergistically aid in the reduction of $Ni^{2+}$ and facilitate the dissolution of Ni atoms into the alloy.

With greater Ni content (between ~35% and ~65%), Ni species can no longer be fully reduced to form an alloy with Cu by annealing at 500 °C in a reasonable time scale (up to 24 h). Instead, heterojunctions between an oxide with lattices close to NiO, and a metallic domain with lattices resembling Cu, were identified by correlated EDS mapping and HRTEM (Figure 3C and Figure S14). The Moiré fringes at the junction provide further evidence of distinct lattices on each side of the boundary. Notably, the contact area between the oxide and metallic regions is small, compared to the broad interfaces observed in the Cu-Ni-O oxide-oxide heterojunctions. This observation indicates that there is a high interfacial energy between the metal and oxide. Therefore, the high-energy interface is minimized as each component forms a semi-spherical morphology. With even greater Ni content (> ~65%), rock salt-structured Cu-Ni-O mixed oxide NPs were formed that do not differ significantly from the products obtained at lower temperatures (300 °C).

[Scheme 1, single column]

The observations reported above show the structural fates of NPs in a homogeneous (Co-Ni-O) and an inhomogeneous system (Cu-Ni-O). The inhomogeneous system, unlike the homogeneous one, can form multiple phase-segregated products depending on composition and temperature, as summarized in Scheme 1. This knowledge can inform the synthesis of more complicated systems like quaternary Co-Ni-Cu-O NPs. To demonstrate how this system behaves, arrays of these NPs were generated by patterning polymer nanoreactors containing Co, Ni, and Cu nitrate precursors, followed by annealing at 500 °C under Ar (Figure S15A). In this system, the Co and Ni are primarily localized in the same domain, as predicted by our two-component study. With increasing Cu content, the NPs exhibited a homogeneous–phase-segregated–homogeneous particle structural evolution (Figure S15B), similar to what we have observed in the Cu-Ni-O system at 500 °C.

In addition to being the first example of using polymer nanoreactor-mediated synthesis to prepare multicomponent oxide NPs with homogenous and phase-segregated structures, this work provides important insight into the environmental and stoichiometric factors that govern particle architecture. Significantly, depending on the ratio of the metal components, a reduced metal component (Cu) can facilitate the reduction of other metal elements (Ni), which typically form stable oxides in isolated NPs. In addition, we have shown that the behavior of higher ordered metal-oxides (three metal components) is similar to that observed for lower ordered systems (two components), suggesting that lower ordered systems can be blueprints for designing higher ordered structures. Note that we are not restricted to 3 metal elements,[11b] as the work herein is just proof-of-concept. Moreover, these complex nanostructures can be readily generated en masse with large area lithographic techniques such as polymer pen lithography.[7a, 18] Taken together, all of these observations should lead to an enhanced understanding of the conditions that yield complex oxide NPs with emergent properties.

## Acknowledgement

This material is based upon work supported by GlaxoSmithKline LLC; the Air Force Office of Scientific Research awards FA9550-16-1-0150, FA9550-12-1-0141, FA9550-12-1-0280; the U.S. Army grant W911NF-15-1-0151; the National Science Foundation award DMR-1507810; and the National Cancer Institute of the National Institutes of Health award U54CA199091. The content is solely the responsibility of the authors and does not necessarily represent the official views of the NIH. J.S.D. acknowledges support from a Hierarchical Materials Cluster Program Fellowship from Northwestern University. P.-C.C. acknowledges support from a Ryan Fellowship from Northwestern University. This work made use of the EPIC Facilities of the NU*ANCE* Center supported by the SHyNE Resource NNCI site (NSF ECCS-1542205), the MRSEC program (NSF DMR-1121262), the IIN, and the State of Illinois through the IIN; the J. B. Cohen X-Ray Diffraction Facility supported by MRSEC and SHyNE; and the Structural Biology Facility supported by NCI CCSG P30 CA060553 awarded to the Robert H Lurie Comprehensive Cancer Center, and the Chicago Biomedical Consortium with support from the Searle Funds at The Chicago Community Trust.

**Keywords:** structural evolution • mixed oxide • doping • nanoparticles • scanning probe lithography

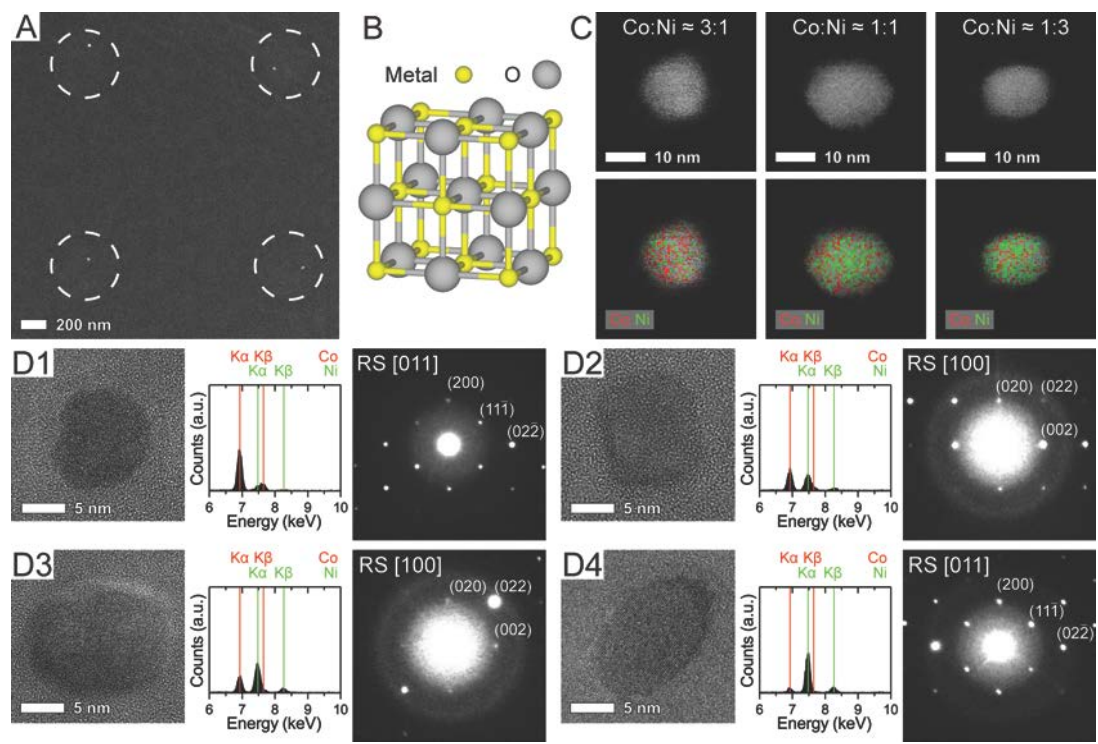

**Figure 1.** (A) HAADF image of an array of Ni-rich oxide NPs. White dashes indicate the original location of individual polymer nanoreactors. (B) The crystal structure of a rock salt (RS) mixed oxide. (C) HAADF imaging (first row) and corresponding EDS mapping (second row) results for Co-Ni-O NPs with different metal contents. (D) Correlated HRTEM (left), EDS (middle), and electron diffraction (right) results for Co-Ni-O NPs with different metal contents with a rock salt (RS) structure. Ni content: (D1) 8%, (D2) 40%, (D3) 60%, (D4) 89%.

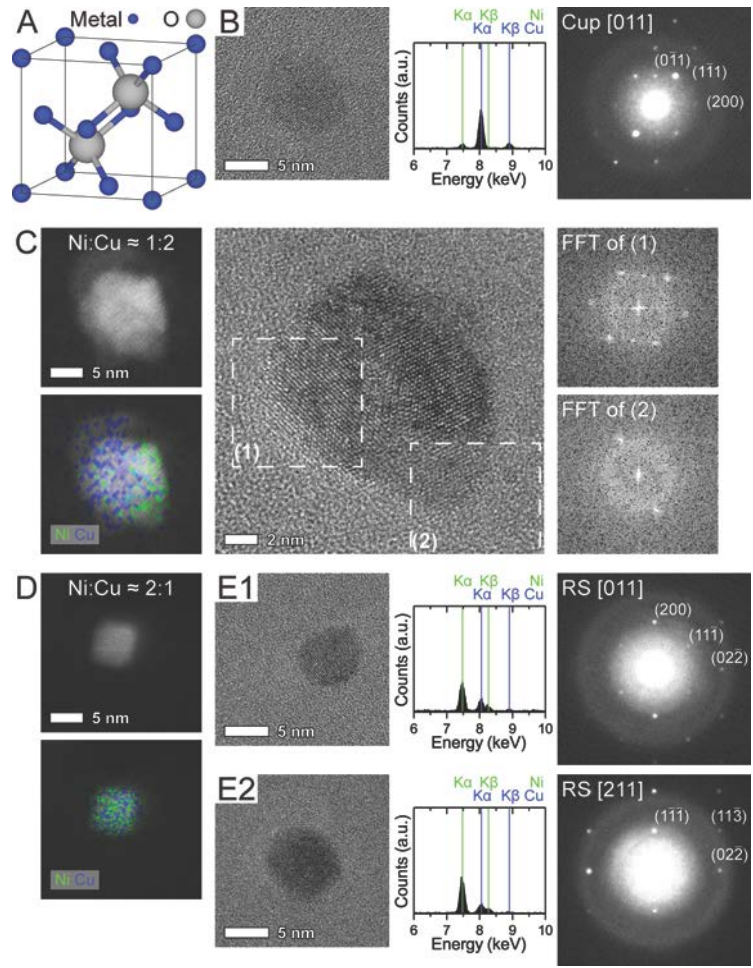

**Figure 2.** (A) Crystal structure of cuprite (e.g., $Cu_2O$). (B) Correlated HRTEM (left), EDS (middle), and electron diffraction (right) results for a Cu-rich Cu-Ni-O NP with a cuprite (Cup) structure. Ni content: 11%. (C) Correlated HAADF imaging, EDS mapping (first column), HRTEM (second column), and FFT lattice analysis (third column) results for a Cu-Ni-O oxide-oxide heterojunction with an intermediate Ni content of ~33%. (D) HAADF imaging and corresponding EDS mapping results for a Ni-rich Cu-Ni-O NP. (E) Correlated HRTEM (left), EDS (middle), and electron diffraction (right) results for Ni-rich Cu-Ni-O NPs with a rock salt (RS) structure. Ni content: (E1) 74%, (E2) 82%.

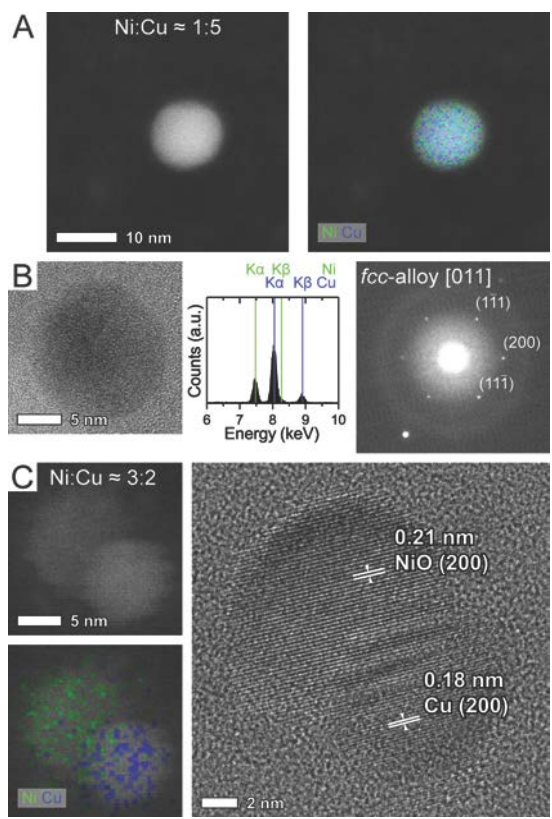

**Figure 3.** (A) HAADF imaging (left) and corresponding EDS mapping (right) results of a Cu-rich Cu-Ni metallic NP. (B) Correlated HRTEM (left), EDS (middle) and electron diffraction (right) results for a Cu-rich Cu-Ni metallic NP. Ni content: 32%. (C) Correlated HAADF imaging, EDS mapping (left column) and HRTEM (right column) results for a Cu-Ni-O metal-oxide heterojunction.

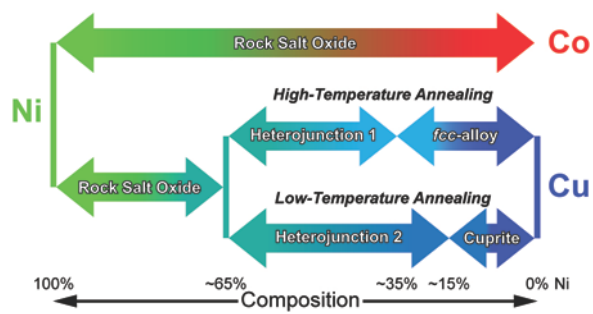

**Scheme 1.** A summary of the crystal structure and architecture of bimetallic hybrid metal-oxide NPs with different compositions of Ni combined with either Co or Cu. For particles containing Ni and Co, all compositional ratios and annealing conditions yield a rock salt oxide. For particles containing Ni and Cu, when the amount of Ni is higher than ~65%, a rock salt structure is obtained. Below ~65% but above ~35% Ni under high temperature annealing (500 °C) a metal-oxide heterojunction structure is formed (Heterojunction 1). At lower annealing temperatures (300 °C) and over a slightly broader compositional range (~65% to ~15% Ni) an oxide-oxide junction forms (Heterojunction 2). At higher Cu contents an *fcc*-alloy forms under high annealing temperatures while a cuprite structure forms under low temperature conditions.

**Entry for the Table of Contents**

# COMMUNICATION

**Deciphering the fate of complex oxide nanoparticles**: A series of hybrid nanoparticles consisting of the elements, Co, Ni, Cu, and O, were synthesized in polymer nanoreactors and investigated by correlated electron microscopy. The structures of these nanoparticles, either mixed oxides, alloys, or heterojunctions of oxide-oxide, or metal-oxide are jointly determined by their miscibility, composition, and annealing temperature.

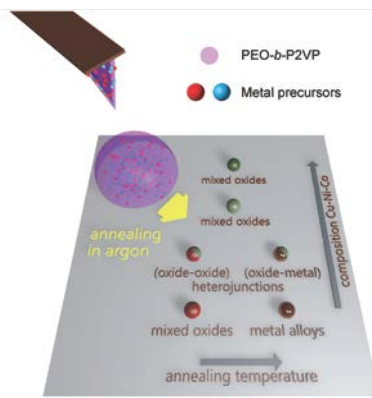

*J. S. Du, P.-C. Chen, B. Meckes, Z. Xie, J. Zhu, Y. Liu, V. P. Dravid,\* C. A. Mirkin\**

*Page No. – Page No.*

**The Structural Fate of Individual Multicomponent Metal-Oxide Nanoparticles in Polymer Nanoreactors**